\ifx\pdfoutput\undefined
\documentclass[aps,prd,a4paper,twoside,twocolumn,english,superscriptaddress,
	showkeys,nofootinbib,showpacs,
	amssymb,amsmath,amsfonts]{revtex4}
\else
\documentclass[aps,prd,a4paper,twoside,twocolumn,english,superscriptaddress,
	showkeys,nofootinbib,showpacs,
	amssymb,amsmath,amsfonts]{revtex4}

\fi

\usepackage{babel}
\usepackage{amsmath}
\usepackage{amsfonts}
\usepackage{amssymb}
\usepackage{dsfont}

\begin{document}

\setcounter{page}{1}

\title{From the Einstein-Cartan to the Ashtekar-Barbero canonical constraints, passing through the Nieh-Yan functional}

\author{\firstname{Simone} \surname{Mercuri}}
\email{mercuri@cpt.univ-mrs.fr}
\affiliation{Centre de Physique Th\'eorique de Luminy, Universit\'e de la M\'editerran\'ee, F-13288, Marseilles, EU;}
\affiliation{Dipartimento di Fisica, Universit\`a di Roma ``La Sapienza'', P.le Aldo Moro 5, I-00185, Rome, EU.}

\begin{abstract}
The Ashtekar-Barbero constraints for General Relativity with fermions are derived from the Einstein-Cartan canonical theory rescaling the state functional of the gravity-spinor coupled system by the exponential of the Nieh-Yan functional. A one parameter quantization ambiguity naturally appears and can be associated with the Immirzi parameter.
\end{abstract}



\maketitle

\section{Introduction}\label{par1}

The growing interest that the Ashtekar formulation of General Relativity (GR) \cite{Ash1986-87} has gained over the years is a consequence of the remarkable simplification it introduces in the structure of the canonical constraints. Specifically, by using the self-dual $SL(2,\mathbb{C})$ Ashtekar connections as fundamental variables, the constraints of GR reduce to a polynomial form. On the other hand, the complex character of the self-dual connections forces to introduce suitable reality conditions, which, until now, have prevented the construction of a complete quantum theory. This led to the adoption of the real Ashtekar-Barbero $SU(2)$ or $SO(3)$ valued connections as fundamental variables in a partially gauge fixed version of tetrads GR \cite{Ash1987-88,Bar1995}. By using the new variables Ashtekar, Rovelli and Smolin constructed a non-perturbative (background independent) quantum theory of gravity known as \emph{Loop Quantum Gravity} (LQG) \cite{AshRovSmo1992,AshLew2004,Rov2004,Thi2001}, which, at least in symmetric physical systems, allows to cure the inevitable singular behavior of General Relativity \cite{Mod2004,AshPawSin2006,AshBoj2006,BojHerKag2006}. The results recently obtained about the graviton propagator have further strengthened the theory, by providing other evidences on its non-singular behavior \cite{Rov2006,BiaModRov2006,ChrLivSpe2007}. 

The Ashtekar-Barbero connections contain a free parameter in their definition, called the Immirzi parameter and here denoted as $\beta$. The presence of the Immirzi parameter does not affect the classical theory, but it appears in the spectra of geometrical non-perturbative quantum operators as, for instance, the Area and Volume operators. Even though many attempts have been made to understand the physical meaning of the Immirzi parameter \cite{RovThi1998,GamObrPul1999,GarMar2003,PerRov2005,FreMinTak2005,ChoTunYu2005}, its role in canonical quantum gravity has not completely clarified yet \cite{Mer2007-1}.

In this brief paper, motivated by the interesting analysis proposed by Gambini, Obregon and Pullin \cite{GamObrPul1999} and by using the results of previous works \cite{Mer2006-1,Mer2006-2}, we will be demonstrating that the Ashtekar-Barbero constraints can be derived starting from the canonical Einstein-Cartan theory, by considering a topologically suggested rescaling of the wave function describing the quantum states of the system. This demonstration provides us with an interesting hint for interpreting the Immirzi parameter which, as the so called $\theta$-angle of QCD, is a quantization ambiguity connected with the non-trivial structure of the quantum configuration space \cite{Mer2007-1}.

The paper is organized as follows: in Section \ref{par2} we briefly recall the canonical formulation of gravity and write the constraints of the Einstein-Cartan action. Then we recall the expression of the non-minimal action introduced in \cite{Mer2006-1} (see also the interesting paper \cite{Ran2005}). Analogously to the Holst action, the non-minimal action deviates from the Einstein-Cartan, specifically it is characterized by two modifications, which, as soon as the solution of the second Cartan structure equation is taken into account, reduce to the Nieh-Yan topological term \cite{Mer2006-1,NieYan1982}. In Section \ref{par3} we are going to demonstrate that this fact makes it possible to calculate the Ashtekar-Barbero constraints for GR starting from the canonical constraints of the Einstein-Cartan theory in the time gauge by using an original and interesting method, which sheds some light on the underlying topological structure of the theory. Namely, after having canonically quantized the Einstein-Cartan theory, we rescale the state functional by the exponential of the Nieh-Yan functional, which here plays a role analogous to the one that the Chern-Simons functionals play in Yang-Mills gauge theories (the Immirzi parameter being analogous to the angular parameter usually indicated by $\theta$) and demonstrate that the canonical constraints of the non-minimal action can be easily derived once the modification to the conjugated variables are calculated. Finally, in Section \ref{par4} we discuss the results obtained.

The signature throughout the paper is $+,-,-,-$. We assume $8\pi G=1$.

\section{Canonical GR with fermion fields}\label{par2}

The interaction between gravity and fermion fields is described by the Einstein-Cartan action,
\begin{align}\label{Einstein-Cartan action}
S\left(e,\omega,\psi,\overline{\psi}\right)&=\frac{1}{2}\int e_{a}\wedge e_{b}\wedge\star\,R^{a b}
\\\nonumber
&+\frac{i}{2}\int\star\; e_a\wedge\left[\overline{\psi}\gamma^a \mathcal{D}\psi-\overline{\mathcal{D}\psi}\gamma^a\psi
\right]\,.
\end{align}
The covariant derivatives contain the Lorentz valued spin connections according to the following definitions:
\begin{equation}\label{spinor cov der}
\mathcal{D}\psi=d\psi-\frac{i}{4}\,\omega^{a b}\Sigma_{a b}\psi\quad\text{and}\quad
\overline{\mathcal{D}\psi}=d\overline{\psi}+\frac{i}{4}\,\overline{\psi}\Sigma_{a
b}\omega^{a b}\,,
\end{equation}
with $a,b,c\dots=0,\dots,3$. Assuming that space-time is globally hyperbolic, we can extract the $3+1$ form of the above action and, once the temporal gauge is fixed and the second class constraints solved \cite{BarSa2000} (further details about this procedure in the presence of spinor fields will be given in \cite{Mer2007-1}), we can extract the following canonical constraints:\footnote{It is worth recalling that the additional strong equation $D_{\alpha}E^{\beta}_i=0$ follows from the solution of the second class constraints, representing the so called \emph{compatibility condition}.}
\begin{subequations}\label{reduced constraints}
\begin{align}
\mathcal{R}_i&:=\epsilon_{i j}^{\ \ k}K_{\alpha}^j E^{\alpha}_k -\frac{i}{4}\,\overline{\Pi}\Sigma_i\psi+\frac{i}{4}\,\overline{\psi}\Sigma_i\Omega\approx 0\,, 
\\
\mathcal{C}_{\alpha}&:=2E^{\beta}_i \mathcal{D}_{[\alpha}K_{\beta]}^i +\overline{\Pi}\mathcal{D}_{\alpha}\psi+\overline{\mathcal{D}_{\alpha}\psi}\Omega\approx 0\,,
\\\label{scalar constraint}
\mathcal{C}&:=-\frac{1}{2}E^{\alpha}_i E^{\beta}_{k}\left(^{(3)}R_{\alpha\beta}^{\ \ \ i k}-2K_{[\alpha}^i K_{\beta]}^k\right)
\\\nonumber
&\ \ \ \ -i E^{\alpha}_{i}\left(\overline{\Pi}\Sigma^{0i}\mathcal{D}_{\alpha}\psi-\overline{\mathcal{D}_{\alpha}\psi}\Sigma^{0i}\Omega\right)
\\\nonumber
&\ \ \ \ +\frac{i}{2}\,\epsilon^{i}_{\ k l}E^{\alpha}_{i}K_{\alpha}^k\left(\overline{\Pi}\Sigma^l\psi-\overline{\psi}\Sigma^l\Omega\right)
\\\nonumber
&\ \ \ \ +\frac{1}{2}E^{\alpha}_{k}K_{\alpha}^k\left(\overline{\Pi}\psi+\overline{\psi}\Omega\right)\approx 0
\,,
\end{align}
\end{subequations}
where we have introduced the following notations: Greek indexes indicate the spatial components of space-time tensors $\alpha,\beta,\dots=1,2,3$, while Latin indexes refer to the internal degrees of freedom $i,j,\dots=1,2,3$; $E^{\alpha}_i=-e\,e^{\alpha}_i$ and $K_{\beta}^j=\omega_{\beta}^{0j}$ form a canonical pair and represent respectively the densitized triad and the the extrinsic curvature; $\overline{\Pi}=\frac{i}{2}\,e\overline{\psi}\gamma^0$ and $\Omega=-\frac{i}{2}\,e\gamma^0\psi$ are respectively the momenta conjugate to $\psi$ and $\overline{\psi}$; $\Sigma_i=\frac{1}{2}\,\epsilon_{i}^{\ j k}\Sigma_{j k}$ are the generators of spatial rotations. Moreover the new derivative symbol, $\mathcal{D}$, representing the connections of the rotations group, has been introduced. We would like to stress that the apparent doubling of the fermionic degrees of freedom is a consequence of the specific form of the action we started from. The right number of degrees of freedom can be restored simply imposing a suitable set of second class constraints, which make it possible to refer the dynamics of the fermion sector of the theory to only one of the two pairs of fermion conjugate fields: either $\left(\psi,\overline{\Pi}\right)$ or $\left(\Omega,\overline{\psi}\right)$.

Once the phase space is equipped with the following symplectic structure
\begin{subequations}
\begin{align} \left\{K^{i}_{\alpha}(t,x),E^{\beta}_{k}(t,x^{\prime})\right\}&=\delta_{\alpha}^{\beta}\delta^{i}_{k}\delta^{(3)}(x-x^{\prime})
\\
\left\{\psi^A(t,x),\overline{\Pi}_B(t,x^{\prime})\right\}_{+}&=\delta^{A}_{B}\delta^{(3)}(x-x^{\prime})\,,
\\
\left\{\Omega^B(t,x^{\prime}),\overline{\psi}_A(t,x)\right\}_{+}&=\delta_{A}^{B}\delta^{(3)}(x-x^{\prime})\,,
\end{align}
\end{subequations}
we can verify that the constraints (\ref{reduced constraints}) are first class. They are connected with the gauge freedom of the theory, namely invariance under spatial rotations and space-time diffeomorphisms\footnote{The theory was initially invariant under the full Lorentz group, but, as well known, the temporal gauge fixes the boost sector, leaving a remnant invariance under the local spatial rotations only.}. 

In \cite{Mer2006-1} we demonstrated that the non-minimal action
\begin{align}\label{new action}
S\left(e,\omega,\psi,\overline{\psi}\right)&=\frac{1}{2}\int e_{a}\wedge e_{b}\wedge\left(\star\,R^{a b}-\frac{1}{\beta}\,R^{a b}\right)
\\\nonumber
&+\frac{i}{2}\int\star\; e_a\wedge\left[\overline{\psi}\gamma^a\left(1-\frac{i}{\beta}\,\gamma_5\right) \mathcal{D}\psi+h.c.
\right]
\end{align}
has the following characteristics: i) it has the Einstein-Cartan action without any modification as effective limit, consequently the Immirzi parameter disappears from the classical equations as in the original (purely gravitational) Holst approach; ii) it can be extended to arbitrary complex values of the Immirzi parameter, in particular for $\beta=\pm\,i$ it reduces to the Ashtekar-Romano-Tate action \cite{AshRomTat1989}; iii) it is suitable for a geometrical description, since the non-minimal term in the fermionic sector together with the Holst modification can be reduced to the Nieh-Yan invariant \cite{NieYan1982} once the second Cartan structure equation is solved. In other words, the modifications introduced in (\ref{new action}) with respect to the Einstein-Cartan action (\ref{Einstein-Cartan action}) can be reduced to the Nieh-Yan invariant, i.e.
\begin{align}\label{nieh-yan}
\nonumber&\frac{1}{2\beta}\int\left[R^{a b}\wedge e_{a}\wedge e_{b}+\star\, e_a\wedge\left(\overline{\psi}\gamma^5\gamma^a\mathcal{D}\psi-\overline{\mathcal{D}\psi}\gamma^a\gamma^5\psi\right)\right]
\\\nonumber
&=\frac{1}{2\beta}\left[\int\left(R^{a b}\wedge e_{a}\wedge e_{b}-T^a\wedge T_a\right)-\int d\left(\star\,e_a J^a_{(A)}\right)\right]
\\
&=\frac{1}{2\beta}\int d\left(e_a\wedge T^a\right)\,,
\end{align}
where $J^d_{(A)}=\overline{\psi}\gamma^d\gamma^5\psi$. Passing from the first to the second line, we used the solution of the Cartan structure equation, namely
\begin{equation}\label{solution}
\omega^{a b}=e^a_{\ \rho}\nabla_{\mu} e^{\rho b} dx^{\mu}+\frac{1}{4}\,\epsilon^{ab}_{\ \ c d}e^c J^d_{(A)}\,,
\end{equation}
resulting from the variation of the action (\ref{new action}) with respect to the spin connection, which, once the factor $\frac{1}{2}\,\epsilon^{a b}_{\ \ c d}-\frac{1}{\beta}\,\delta^{[a}_{c}\delta^{b]}_{d}$ is dropped, is equivalent to the one characterizing the usual Einstein-Cartan theory. Finally, the result in the third line can be easily obtained using the following identity $T^a\wedge e_a=-\frac{1}{2}\star\,e_a J^a_{(A)}$. 
So, the action (\ref{new action}) is dynamically equivalent to the usual Einstein-Cartan action, since the additional terms introduced in action (\ref{new action}) with respect to (\ref{Einstein-Cartan action}) can be reduced to a total divergence containing torsion. The fact that the effective action derived from (\ref{Einstein-Cartan action}) contains a topological modification with respect to (\ref{new action}) suggests that the two sets of canonical constraints resulting from them have to be connected by a topologically motivated redefinition of the conjugate momenta. In fact, in the next section we are going to derive the constraints of the non-minimal theory (\ref{new action}) by redefining the state functional of the formally quantized Einstein-Cartan theory, namely by rescaling it by the exponential of the Nieh-Yan functional
\begin{equation} 
\mathcal{Y}(e,\psi,\overline{\psi})=\int e_i\wedge T^i(e,\psi,\overline{\psi})\,,
\end{equation}
where we have to take into account the solution of the second Cartan structure equation,
which for spatial internal indexes gives
\begin{equation}\label{torsion}
T^i(e,\psi,\overline{\psi})=\frac{1}{2}\,\epsilon^i_{\ j k}e^{j}\wedge e^k J^0_{(A)}\,.
\end{equation}

\section{From Einstein-Cartan to Ashtekar-Barbero constraints}\label{par3}

Weak equations (\ref{reduced constraints}) represent a set of first class constraints, so we can quantize the system by adopting the Dirac procedure, i.e. the constraints are directly implemented in the quantum theory by requiring that the wave functional be annihilated by their operator representation. Let me assume as coordinates on the quantum configuration space the following fields: $E^{\alpha}_i$, $\psi^A$ and $\overline{\psi}_B$, so that the wave function is 
\begin{equation}
\Phi=\Phi\left(E,\psi,\overline{\psi}\right)\,.
\end{equation}
The quantum gravitational equations are formally expressed as
\begin{subequations}
\begin{align}
\widehat{\mathcal{R}}_i\Phi\left(E,\psi,\overline{\psi}\right)=0\,,
\\
\widehat{\mathcal{C}}_{\alpha}\Phi\left(E,\psi,\overline{\psi}\right)=0\,,
\\
\widehat{\mathcal{C}}\,\Phi\left(E,\psi,\overline{\psi}\right)=0\,,
\end{align}
\end{subequations}
where the operatorial translation of the canonical constraints relies on the following prescriptions
\begin{subequations}
\begin{align}
\widehat{E}^{\alpha}_i\Phi\left(E,\psi\right)&=E^{\alpha}_i\Phi\left(E,\psi\right)\,,\qquad
\\
\widehat{K}_{\beta}^k\Phi\left(E,\psi\right)&=i\frac{\delta}{\delta E^{\beta}_k}\Phi\left(E,\psi\right)\,,
\end{align}
\end{subequations}
\begin{subequations}
\begin{align}
\widehat{\psi}^A\Phi\left(E,\psi\right)&=\psi^A\Phi\left(E,\psi\right)\,,
\\
\widehat{\overline{\Pi}}_B\Phi\left(E,\psi\right)&=-i\frac{\delta}{\delta\psi^B}\Phi\left(E,\psi\right)\,.
\end{align}
\end{subequations}

Suppose now to rescale the state functional $\Phi\left(E,\psi\right)$ by the exponential of the Nieh-Yan functional\footnote{For a concise explanation of this procedure for $SU(N)$ Yang-Mills gauge theories see \cite{AshBalJo1989}.} 
\begin{equation}\label{rescaling}
\Phi^{\prime}(E,\psi,\overline{\psi})=\exp\left\{\frac{i}{\beta}\,\mathcal{Y}(E,\psi,\overline{\psi})\right\}\Phi(E,\psi,\overline{\psi})\,,
\end{equation}
where the parameter $\beta$ will result to be associated with the Immirzi parameter.
It is worth stressing that the study of large gauge transformations in temporal gauge fixed gravity leads precisely to the rescaling (\ref{rescaling}) of the state functional, so a consistent geometrical interpretation of the entire procedure can be provided, but this discussion is beyond the scope of this brief letter and will be presented in a longer, forthcoming paper \cite{Mer2007-1}. 

The price paid for this rescaling is a modification in the definitions of the canonical conjugate momentum operators $\widehat{K}^i_{\alpha}$, $\widehat{\overline{\Pi}}_B$ and $\widehat{\Omega}^B$. Specifically we get: 
\begin{subequations}
\begin{align}
\nonumber&\widehat{K}^{\prime\,i}_{\ \alpha}\Phi^{\prime}(E,\psi,\overline{\psi})=e^{\frac{i}{\beta}\,\mathcal{Y}(E,\psi,\overline{\psi})}\widehat{K}^{i}_{\ \alpha}e^{-\frac{i}{\beta}\,\mathcal{Y}(E,\psi,\overline{\psi})}\Phi^{\prime}(E,\psi,\overline{\psi})
\\
&\ \ \ \ \ =\left(\widehat{K}^i_{\alpha}-\frac{1}{2\beta}\,\epsilon^{\ \beta\gamma}_{\alpha}\widehat{T}^{i}_{\beta\gamma}\right)\Phi^{\prime}(E,\psi,\overline{\psi})\,,
\\
\nonumber&\widehat{\overline{\Pi}}\,^{\prime}_B\Phi^{\prime}(E,\psi,\overline{\psi})=e^{\frac{i}{\beta}\,\mathcal{Y}(E,\psi,\overline{\psi})}\widehat{\overline{\Pi}}_B e^{-\frac{i}{\beta}\,\mathcal{Y}(E,\psi,\overline{\psi})}\Phi^{\prime}(E,\psi,\overline{\psi})
\\
&\ \ \ \ \ =\left(\widehat{\overline{\Pi}}_B-\frac{i}{\beta}\widehat{\overline{\Pi}}_A\left(\gamma^5\right)^A_{\ B}\right)\Phi^{\prime}(E,\psi,\overline{\psi})\,,
\\
\nonumber&\widehat{\Omega}^{\prime\,B}\Phi^{\prime}(E,\psi,\overline{\psi})=e^{\frac{i}{\beta}\,\mathcal{Y}(E,\psi,\overline{\psi})}\widehat{\Omega}^{B}e^{-\frac{i}{\beta}\,\mathcal{Y}(E,\psi,\overline{\psi})}\Phi^{\prime}(E,\psi,\overline{\psi})
\\
&\ \ \ \ \ =\left(\widehat{\Omega}^{B}-\frac{i}{\beta}\left(\gamma^5\right)^{B}_{\ A}\widehat{\Omega}^{A}\right)\Phi^{\prime}(E,\psi,\overline{\psi})\,,
\end{align}
\end{subequations}
where we have formally calculated the functional derivatives, taking into account expression (\ref{torsion}) and the following useful form of the Nieh-Yan functional
\begin{equation}
\mathcal{Y}(E,\psi,\overline{\psi})=-\frac{1}{2}\int d^3x\,\epsilon_{\gamma}^{\ \alpha \beta}E^{\gamma}_i  T_{\alpha\beta}^i\,.
\end{equation}
The modifications of the conjugate momentum operators correspond to canonical modifications of the respective classical conjugate momenta, which can be easily evaluated. So that, reintroducing the new classical conjugate momenta into the canonical constraints (\ref{reduced constraints}) we obtain the following weak equations:
\begin{subequations}\label{new constraints}
\begin{align}
\mathcal{R}_i^{\prime}&=\frac{1}{\beta}\mathcal{D}_{\alpha}E^{\alpha}_i+\epsilon_{i j}^{\ \ k}K_{\alpha}^j E^{\alpha}_k
\\\nonumber
&-\frac{i}{4}\,\overline{\Pi}f^5_{(\beta)}\Sigma_i\psi+\frac{i}{4}\,\overline{\psi}\Sigma_if^5_{(\beta)}\Omega\approx 0\,,
\\
\mathcal{C}_{\alpha}^{\prime}&=2E^{\beta}_i\mathcal{D}_{[\alpha}K_{\beta]}^i+\frac{1}{2\beta}\,\epsilon^{i j}_{\ \ k}E^{\beta}_i R_{\alpha\beta j}^{\ \ \ \ k}
\\\nonumber
&+\overline{\Pi}f^5_{(\beta)}\mathcal{D}_{\alpha}\psi+\overline{\mathcal{D}_{\alpha}\psi}f^5_{(\beta)}\Omega\approx 0\,,
\end{align}
\begin{align}
\mathcal{C}^{\prime}&=\frac{1}{2}E^{\alpha}_i E^{\beta}_{k}\left(\epsilon^{i k}_{\ \ l}R_{\alpha\beta}^{l}+2K_{[\alpha}^i K_{\beta]}^k\right)
\\\nonumber
&-i E^{\alpha}_{i}\left(\overline{\Pi}f^5_{(\beta)}\Sigma^{0i}\mathcal{D}_{\alpha}\psi-\overline{\mathcal{D}_{\alpha}\psi}\Sigma^{0i}f^5_{(\beta)}\Omega\right)
\\\nonumber
&+\frac{i}{2}\epsilon^{i}_{\ k l}E^{\alpha}_{i}K_{\alpha}^k\left(\overline{\Pi}f^5_{(\beta)}\Sigma^l\psi-\overline{\psi}\Sigma^l f^5_{(\beta)}\Omega\right)
\\\nonumber
&+\frac{1}{2}E^{\alpha}_{k}K_{\alpha}^k\left(\overline{\Pi}f^5_{(\beta)}\psi+\overline{\psi}f^5_{(\beta)}\Omega\right)\approx 0\,,
\end{align}
\end{subequations}
where $f^5_{(\beta)}=1-\frac{i}{\beta}\gamma^5$, and $R^{l}_{\alpha\beta}=\frac{1}{2}\epsilon^{l}_{\ i k}R_{\alpha\beta}^{\ \ \ i k}$. It is worth noting that the gravitational contributions in the above expressions are exactly those obtainable starting from the Holst action. In fact, the above modified canonical constraints (\ref{new constraints}) can be extracted from a $3+1$ splitting of the non-minimal action (\ref{new action}), so that an interesting topological new aspect of the Ashtekar-Barbero formulation of canonical gravity has been extracted and will be further reinforced by studying the role of the large gauge transformations in this framework \cite{Mer2007-1}.

\section{Discussion}\label{par4}

In this brief paper we have demonstrated that a non-minimal action exists which is dynamically equivalent to the Einstein-Cartan action, since it deviates from the latter due to the presence of two modifications which can easily be reduced to a topological term. This fact suggests a simple analogy with the extension of Yang-Mills gauge theories to contain topological terms. It is well known that in Yang-Mills gauge theories the presence of a local symmetry generates a Gauss constraint in the canonical theory. The Dirac quantization of such a theory requires that the state functional be annihilated by the quantum operator constraint, implying that the state functional has to be invariant under small gauge transformations, i.e. gauge transformations in the connected component of the identity. But the canonical theory does not provide any suggestion on how the state functional behaves under large gauge transformations. The behavior of the wave function under large gauge transformations can be studied by using the Chern-Simons functionals. More specifically, by rescaling the wave function by the exponential of the Chern-Simons we can automatically diagonalize the large gauge transformations operator \cite{GamObrPul1999}. The result is that the modifications introduced by the rescaling affect the Gauss constraint and the Hamiltonian of the theory, providing us with exactly the expression we would have obtained if we started from the well known topological modification of the standard action. 

Here we have demonstrated that the same happens in gravity and, even though the details regarding the large gauge transformations have been omitted to limit the length of the paper (and will be described in \cite{Mer2007-1}), their absence does not prevent one from drawing the conclusion that in temporal gauge fixed gravity the Nieh-Yan functional plays the same role as the Chern-Simons functionals, allowing us to shed some light on the topological aspects of the Ashtekar-Barbero formulation of canonical gravity. 

And, finally, we would like to stress that the Immirzi parameter $\beta$ is introduced in this framework in the rescaling (\ref{rescaling}), where it plays the same role as the so called $\theta$-angle plays in QCD.

\begin{acknowledgments}
The author expresses his gratitude to C. Rovelli for stimulating discussions and suggestions and to the Quantum Gravity group in CPT Marseilles for hospitality. This work is financially supported by ``Fondazione A. Della Riccia''. Shelly Kittleson is also acknowledged for her reading of the manuscript.
\end{acknowledgments}

\end{document}